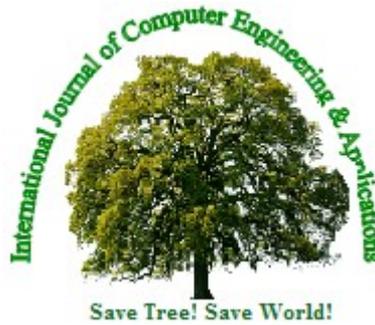

# IMPROVEMENT OF SPECTRUM SHARING USING TRAFFIC PATTERN PREDICTION

**Dr. R. Kaniezhil [1], Dr. C. Chandrasekar [2]**

[1]Department of Computer Applications, Easwari Engineering College, Chennai.

[2]Department of Computer Science, Periyar University, Salem.

**ABSTRACT:**

*The paper focuses on improving the spectrum sharing using NSU, FLS and Traffic Pattern Prediction and also made comparison that traffic pattern prediction provides a better way of improving the spectrum utilization and avoids the spectrum scarcity. This helps to increase the number of active users, ease of identification of optimal users to use the spectrum with maximized coverage of the spectrum.. We experimentally evaluated the effectiveness of our approach using NS2 simulator and showed that after predicting the traffic, we can accommodate more number of users and avoiding Interference.*

## [1] INTRODUCTION

The radio spectrum is a key resource for many new and developing technology based industries. At the same time, it is a vital input into the delivery of many public services. The management and development of the spectrum will therefore play an important role in creating a knowledge driven economy and society. The use of radio spectrum has become an integral part of societies infrastructure. More recently, the phenomenal growth in personal mobile communications has turned wireless access via mobile phones from luxury to necessity for many people. This value to individuals, businesses and the public sector of access to radio spectrum is becoming increasingly recognised. Radio spectrum makes a substantial and increasing contribution to the economy.

Today, Wireless Communications deals with two main problems namely spectrum scarcity and deployment delays. These problems are caused due to the centralized manner of spectrum sharing and static in nature of frequency assignment. This scheme cannot adapt to the changing needs of spectrum by users from the military, government and commercial purposes. New technologies should be used effectively to provide the flexibility needed for the above said problems. Spectrum is no longer sufficiently available, because it has been assigned to the primary users that own the privileges to their assigned spectrum. However, it is not used efficiently most of the time. In order to use the spectrum in an opportunistic manner and to increase the spectrum availability, the unlicensed users can be allowed to utilize the licensed bands of licensed users without causing any interference with the assigned service.

The key enabling technology for dynamic spectrum access techniques is CR networking which allows intelligent spectrum aware devices to opportunistically use the licensed spectrum bands for transmission [1]. CR techniques provide the capability to use or share the spectrum in an opportunistic manner.

## [2] SPECTRUM MANAGEMENT FUNCTIONS

Spectrum management in CRs relates to their ability to rationalize and optimize the use of the radio spectrum. In other words, through spectrum management capabilities they can choose the best possible spectrum band that will meet required QoS amongst all detected available spectrum whitespaces found. The spectrum management functions of a CR include spectrum sensing, spectrum analysis and spectrum decision is shown in the [**Figure-1**]. Spectrum sensing is mainly a physical layer issue while spectrum analysis and decision are more of upper layer issues. In the following sections, descriptions of each of these functions are presented.

The main features of CR are listed as below [2,3,4]
- Spectrum Sensing
- Spectrum Decision
- Spectrum Sharing
- Spectrum Mobility

### 2.1 Spectrum Sensing

It detects the unused spectrum and sharing it without harmful interference with other users. It is an important requirement of the CR network to sense the spectrum holes. Primary users detection is found to be the most efficient way to detect the spectrum holes. Some of the spectrum sensing techniques can be classified as [5] :

#### 2.1.1 Transmitter Detection

In this category, CR must have the capability to determine if a signal from a primary transmitter is locally present in a certain spectrum. Some proposed approaches in this category are Matched Filtering detection, Energy detection, Waveform based sensing, Cyclostationary based sensing, etc.

#### 2.1.2 Cooperative Detection

In this category, it decreases the probabilities of misdetection and false alarm considerably. It can also solve the hidden primary user problem and can decrease the sensing time.

#### 2.1.3 External Detection

In this category, an external agent performs the sensing and broadcasts the channel occupancy information to CR. The main advantage is to overcome the hidden primary user problem as well as the uncertainty due to shadowing and fading. As the CR does not spend time for sensing, spectrum efficiency is increased.

### 2.2 Spectrum Decision

It is the task of capturing the best available spectrum to meet the user requirements. CR should decide on the best spectrum band to meet the QoS requirements over all available spectrum bands, therefore spectrum management functions are required for CRs. This notion is called spectrum decision and constitutes rather important but unexplored topic. Spectrum decision is closely related to the channel characteristics and the operations of primary users.

Spectrum decision usually consists of two steps namely
1. Each spectrum band is characterized based on not only local observations of CR users but also statistical information of primary networks.
2. Based on this characterization, the most appropriate spectrum band can be chosen. This category can be classified as spectrum analysis, spectrum selection and reconfiguration.

#### 2.2.1 Spectrum Analysis

In this technique each spectrum hole should be characterized considering not only the time varying radio environment and but also the primary user activity.

### 2.2.2 Spectrum Selection

When all the analysis of spectrum band is done, appropriate spectrum band should be selected for the current transmission considering the QoS requirements and the spectrum characteristics. According to user requirement the data rate, bandwidth is determined then according to decision rule appropriate spectrum band is chosen.

### 2.2.3 Reconfiguration

The CR users reconfigure communication protocol as well as communication hardware and RF front end according to the radio environment and user QoS requirements.

CR users require spectrum decision in the beginning of the transmission. CR users characterize available spectrum bands by considering the received signal strength, interference and the number of users currently residing in the spectrum which are also used for resource allocation in classical wireless networks.

However, in CR each user observes heterogeneous spectrum availability that is varying over time and space resulting from PU activities. This changing nature of the spectrum usage needs to be considered in the spectrum characterization. Based on this characterization, CR users determine the best available spectrum band to satisfy its QoS requirements. Furthermore, quality degradation of the current transmission can also initiate spectrum decision to maintain the quality of a current session.

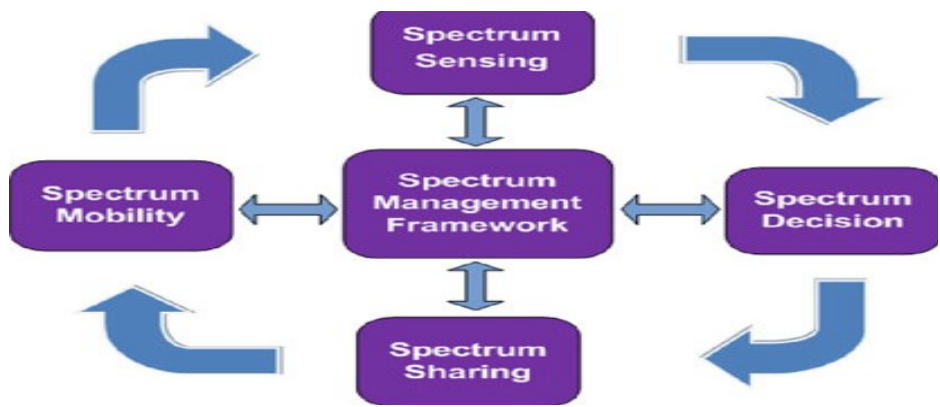

**Figure: 1. Spectrum Management Framework of Cognitive Radio**

### 2.3 Spectrum Sharing

The spectrum access and sharing among licensed and unlicensed users is regulated in a way that the unlicensed or secondary user accesses the spectrum. It should not affect the degree of satisfaction of the licensed users requirements.

It refers to providing the fair spectrum scheduling method, one of the major challenges in the open spectrum usage is the spectrum sharing. CRs have the capability to sense the surrounding environments and allow intended secondary user to increase QoS by opportunistically using the unutilized spectrum holes. If a secondary user senses the available spectrum, it can use this spectrum after the primary licensed user vacates it. Spectrum sharing mainly focuses on resource management within the same spectrum with the following functionalities

### 2.3.1 Resource Allocation

Based on the QoS monitoring results, CR users select the proper channels [6,7,8] and adjust their transmission power [9,10,11] to achieve QoS requirements as well as resource fairness. Especially in power control, sensing results need to be considered as not to violate the interference constraints.

### 2.3.2 Spectrum Access

It enables multiple CR users to share spectrum resources by determining who will access the channel or when a user may access the channel [12,13].

Once a proper spectrum band is selected in spectrum decision, communication channels in that spectrum need to be assigned to a CR user while determining its transmission power to avoid the interference to the primary network. Then, the CR user decides when the spectrum should be accessed to avoid collisions with other CR users.

The infrastructure based network can provide sophisticated spectrum sharing method with support of the Base Station (BS). Thus, it can exploit time slot based scheduling and dynamic channel allocation to maximize the total network capacity as well as achieve fair resource allocation over CR users. Furthermore, through the synchronization in sensing operation the transmission of CR users and primary users can be detected separately. This decouples sensing operation with spectrum sharing.

2.3.3 Spectrum Mobility

It is a process when the CR user exchanges its frequency of operation. CR networks target to use the spectrum in a dynamic manner by allowing the radio terminals to operate in the best available frequency band, maintaining seamless communication requirements during the transition to better spectrum.

In this work, the above said features of CR are applied in order to make spectrum sharing among service providers with the help of CR. Based on these spectrum management functionalities, the secondary users can intelligently change their transmission parameters to accommodate the operating environment. Hence, the secondary users can increase their transmission performance as well as avoid interfering with the primary users.

2.4 Dynamic Spectrum Access

The first step in Dynamic spectrum access is the detection of unused spectral bands. Therefore, CR device is used for measuring the Radio Frequency (RF) energy in a channel to determine whether the channel is idle or not. But, this approach has a problem in that wireless devices can only sense the presence of a primary user (PU) if and only if the energy detected is above a certain threshold. The Taxonomy of Dynamic Spectrum Access is explained using the [**Figure-2**].

Generally, Dynamic spectrum access can be categorized into three models namely [14,15]

    2.4.1 Dynamic Exclusive use model
    2.4.2 Open sharing model
    2.4.3 Hierarchical Access model

2.4.1 Dynamic Exclusive use Model

This model manages spectrum in the finer scale of time, space, frequency and use dimensions. So at any given point in space and time, only one operator has exclusive right to the spectrum but the identity of the owner and type of use can change.

This model maintains the basic structure of the current spectrum regulation policy as spectrum bands are licensed to services for exclusive use. The main idea is to introduce flexibility to improve spectrum efficiency. Two approaches have been proposed under this model are spectrum property rights and dynamic spectrum allocation. The former approach allows licenses to sell and trade spectrum and to freely choose technology.

The other approach, dynamic spectrum allocation aims to improve spectrum efficiency through dynamic spectrum assignment by exploiting the spatial and temporal traffic statistics of different services. Similar to the current static spectrum allotment policy such as strategies allocate at a given time and region, a portion of the spectrum to a radio access network for its exclusive use.

Based on an exclusive use model, these approaches cannot eliminate white space in the spectrum resulting from the bursty nature of wireless traffic.

2.4.2    Open Sharing Model

This model is called as Spectrum commons. This model employs open sharing among peer users as the basis for managing a spectral region. Centralized and Distributed spectrum sharing strategies have been initially investigated to address technological challenges under this model.

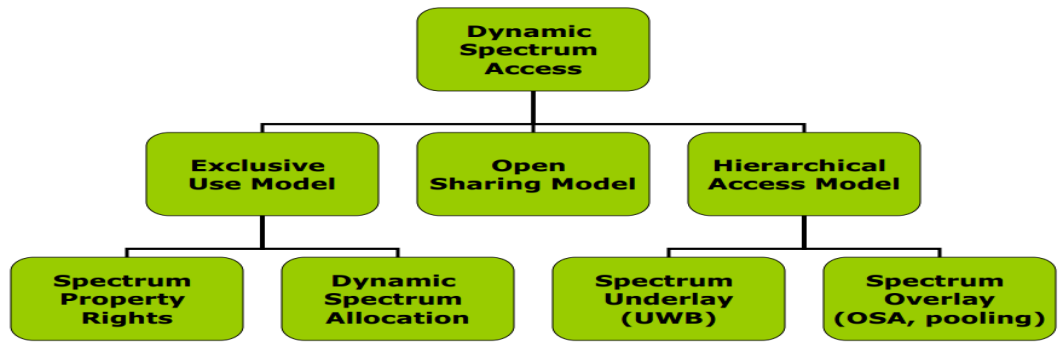

**Figure: 2. Taxonomy of Dynamic Spectrum Access**

2.4.3    Hierarchical Access Model

This model is considered to be the same as an Exclusive use model.  The basic idea is to open licensed spectrum to secondary users and limit the interference perceived by primary users.  This model is first to describe simultaneous shared use of spectrum wherein there is a primary licensed owner of the spectrum band and multiple secondary users opportunistically share the band.  Spectrum sharing between primary and secondary users utilizes spectrum underlay and spectrum overlay approaches.

Sometimes the hierarchical access model can be categorized under the open sharing model because spectrum sharing between primary and secondary users is fundamentally different from spectrum sharing among peer users in both technical and regulatory aspects.

# [3] PROBLEM IN SPECTRUM SHARING

With the advent of new high data rate wireless applications, as well as growth of existing wireless services, demand for additional bandwidth is rapidly increasing.  Government agencies such as FCC allocate frequency spectrum and impose regulations on its usage such as control of features, priorities, allocations and exclusions.  Currently, spectrum allotment operates by providing each new service with its own fixed frequency block.  Demand for access to spectrum has been growing dramatically and is likely to continue to grow in the foreseeable future.

CR technology is a key enabler for both real time spectrum markets and dynamic sharing of licensed spectrum with unlicensed devices.  The CR device is able to perform spectrum acquisition either through purchasing in cleared spectrum or sensing in vacant channels of geographical interleaved spectrum over a range of frequency bands.  In turn, it operates in this spectrum at times and locations where it is able to transmit in a non interfering basis.

The overall spectral efficiency of a system can be improved with good coexistence properties, good spectrum sharing capabilities as well as with flexibility in the spectrum use.  Capabilities to share spectrum with other systems will significantly increase the efficiency as well as acceptability of the system.  The overall spectral efficiency of the work can be also increased with a flexible use of spectrum that adapts to the spatial and temporal variations in the traffic and environment characteristics.

The available spectrum will be utilized opportunistically by the unlicensed users of the licensed spectrum without causing interference to the primary users. Hence, it provides opportunistic spectrum access of using the available spectrum.

Here, the process of spectrum sharing is carried out in long term spectrum assignment methods which access the spectrum for a long period. If the spectrum is accessed in a long term, then there should be a preagreed amount consideration among the service providers.

In this paper, the proposed Cognitive based spectrum access by opportunistic approach of Heterogeneous Wireless networks based on the traffic pattern prediction [16 - 20]. The proposed work made comparison with the Normal Spectrum Utilization (NSU), Fuzzy Logic System (FLS) and with Traffic Pattern Prediction and proved that Spectrum utilization is more using Traffic Pattern Prediction method. So that, there will be more number of active users with minimum Interference.

In this paper, the overall spectrum efficiency is improved by reducing the call blocking rate, interference and by improving the throughput and to keep Quality of Service (QoS) as high as possible. The Spectrum sharing between service providers improves the spectral efficiency, probability efficiency of sensing and reduces the call blockage.

If the number of service providers is increased to share the spectrum, this may reduce the high traffic patterns of the calls. In this thesis, the proposed work presents an approach for evaluating the channel availability and the call arrival rate. The results are used to evaluate the probability of the channel availability of a frequency band within a time period.

## [4] SIMULATION RESULTS

Using NS2 simulation the performance of the overall system efficiency has been evaluated. The main simulation parameters used in this work is shown in Table 1.

In this section, simulation results present the performance of our proposed sensing framework. Channel assignment mechanisms in the traditional multi-channel wireless networks typically select the Best channel for a given transmission. From the proposed work, the available channel with the high probability and high frequency band are chosen.

| Table 1: Simulation Parameters ||
|---|---|
| **Parameters** | **Values** |
| Channel | Wireless Channel |
| Propagation Model | TwoRayGround Model |
| Number of nodes | 150 |
| X Value | 1000 m |
| Y value | 1000 m |
| Number of channels | 4 |
| Number of Base stations | 5 |
| Number of Primary user | 10 |
| Pause time | 12.00 ms |
| Maximum Speed | 2.00 ms |
| Queue length | 100 |
| rxPower | 0.3 J |
| txPower | 0.6 J |
| Service Types | Call Service, Internet service, Multimedia service |

In recent studies, the spectrum allocated by the traditional approach shows that the spectrum allocated to the primary user is underutilized and the demand for accessing the limited spectrum is growing increasingly. Spectrum is no longer sufficiently available, because it has been assigned to the primary users that own the privileges to their assigned spectrum. However, it is not used efficiently most of the time. To increase the spectrum availability, the unlicensed users are allowed to utilize the licensed bands of the licensed users without causing any interference.

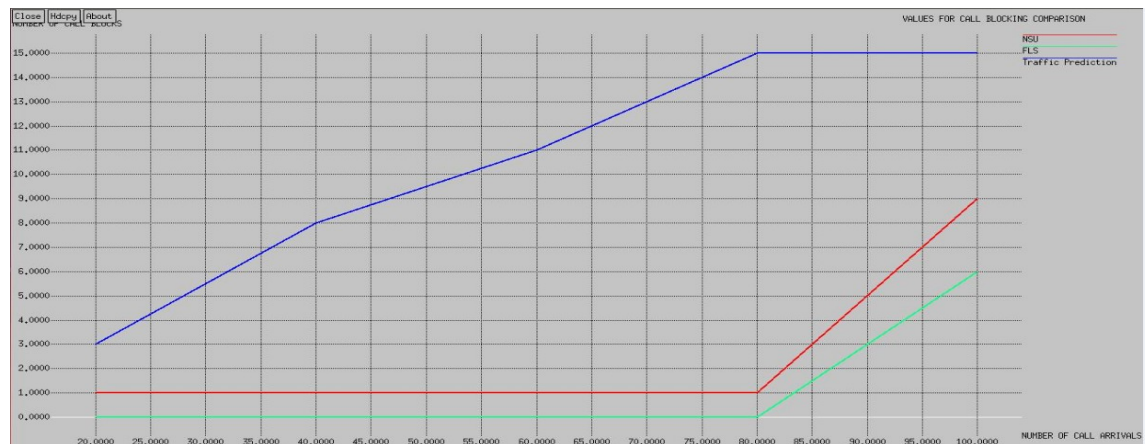

**Figure: 3. Comparison for Call Blocking**

Figure-3 shows the Comparison of call blocking with all the three methods and finally proved that using the traffic pattern prediction has low call blocking which improves the number of active users.

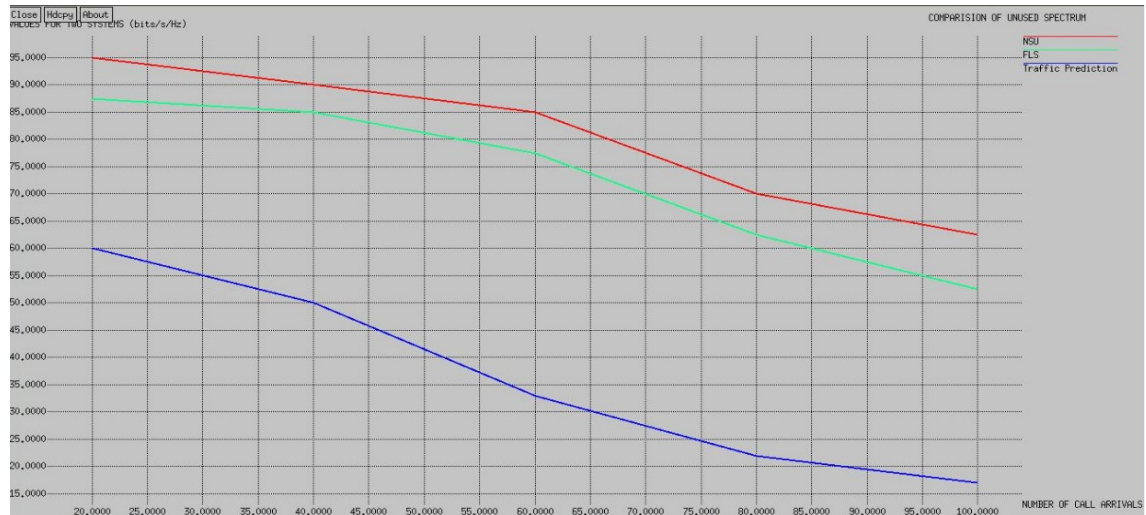

**Figure: 4. Comparison of unused spectrum**

Figure-4 shows the Comparison of unused spectrum which shows that the utilization of spectrum is more in case of traffic pattern. So that, unused spectrum is found to be less in traffic pattern prediction.

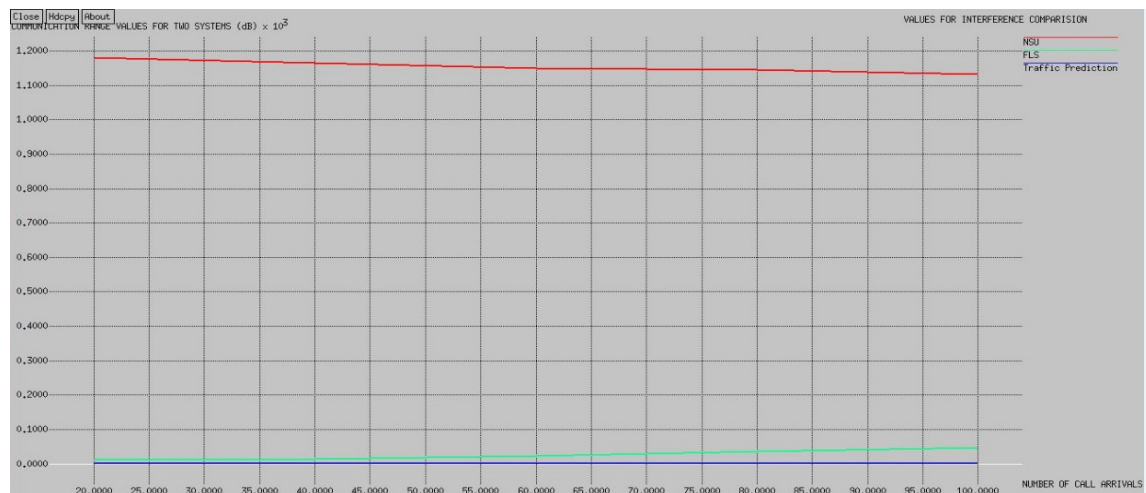
**Figure: 5. Comparison of Interference**

Figure-5 shows the Comparison of Interference in which using NSU and FLS, there is a slight variations in the Interference but using traffic pattern, Interference has been much reduced. So that, there will not be call blocking with the active users.

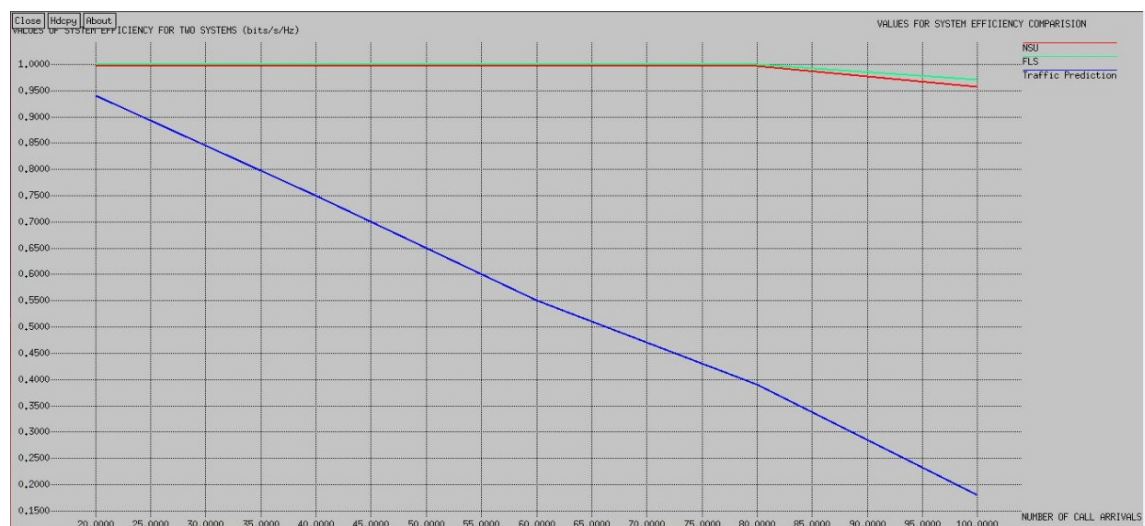
**Figure: 6. Comparison of System Efficiency**

Figure-6 shows the Comparison of system efficiency in which using traffic pattern prediction method system efficiency has been much reduced.

When users of different service providers share the licensed spectrum of the primary user, primary user has the highest priority to access that spectrum. In such a case, the secondary user has to vacate the channel and to move on to another available channel. So that, transmit power will be reduced frequently or communication will be dropped. In order to avoid the temporal connection loss or interference with the primary user, the secondary user has to evaluate the channel availability before using the channels of the primary user and predicting the traffic pattern of the primary user. This would increase the channel utilization is shown in the Figure-7 and reduces the call blockage and interference.

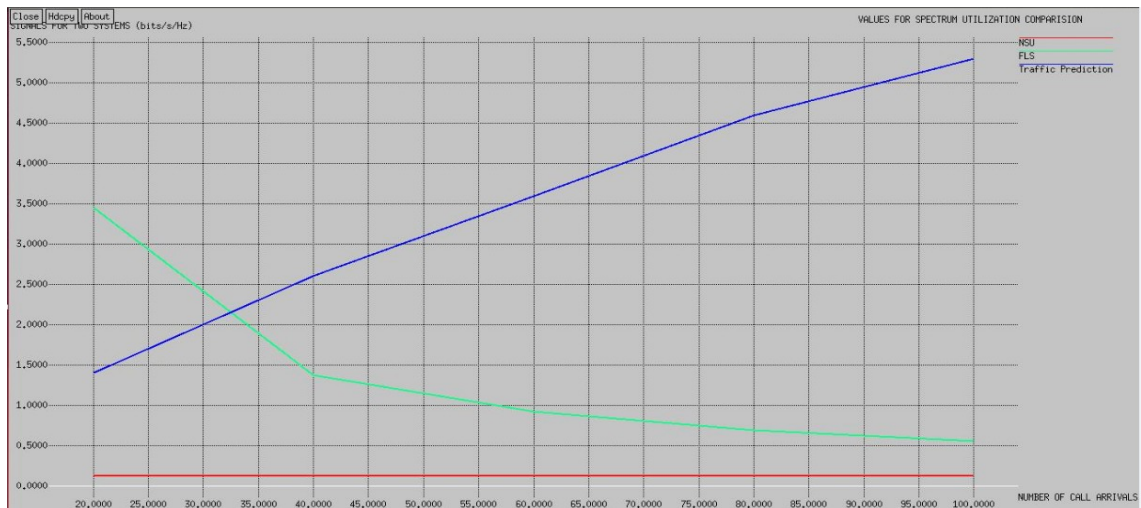
**Figure: 7. Comparison of Spectrum Utilization**

Normal Spectrum Utilization (NSU), FLS and Traffic Prediction Techniques are applied and finally showed that after finding out the traffic prediction, the spectrum utilization is more and it allows the users to utilize the spectrum more efficiently and effectively without scarcity of the spectrum. This leads to the more utilization of the spectrum.

## [6] CONCLUSION

The proposed work was analyzed and simulated using NS2 Simulator in order to validate our approach. This prediction enhances the channel utilization of primary users and enhances the communication of primary users. The simulated results show that there is a decrease in the call blocking and reduced interference. This helps to improve the channel utilization and avoids the scarce resource of the spectrum.